\documentclass[epj]{webofc}
\usepackage[utf8]{inputenc}
\usepackage[varg]{txfonts}   
\usepackage{booktabs}
\usepackage{xcolor}
\usepackage{wrapfig}
\definecolor{darkred}{rgb}{0.4,0.0,0.0}
\definecolor{darkgreen}{rgb}{0.0,0.4,0.0}
\definecolor{darkblue}{rgb}{0.0,0.0,0.4}
\usepackage[bookmarks,linktocpage,colorlinks,
    linkcolor = darkred,
    urlcolor  = darkblue,
    citecolor = darkgreen]{hyperref}
\usepackage{subfigure}
\wocname{EPJ Web of Conferences}
\woctitle{Lattice2017}
\newcommand{\NSigma}{N_\sigma}
\newcommand{\NTau}{N_\tau}

\newcommand{\Nf}{N_\text{f}}

\newcommand{\Loewe}{LOEWE-CSC}
\newcommand{\Lcsc}{L-CSC}

\newcommand{\clqcd}{CL\kern-.25em\textsuperscript{2}QCD}
\newcommand{\Ocl}{OpenCL}
\newcommand{\Amd}{AMD }

\newcommand{\sectionname}{Sec.}

\newcommand{\psibar}{\bar{\psi}} 
\newcommand{\chiralcond}{\langle \psibar \psi \rangle}
\newcommand{\mpi}{m_{\pi}}
\newcommand{\ms}{m_{s}}
\newcommand{\mud}{m_{u,d}}
\newcommand{\Action}{\mathcal S}
\newcommand{\SGluon}{\Action_{\text{G}}}

\newcommand{\LatSpacing}{a}
\newcommand{\LatMassWilson}{\kappa}
\newcommand{\LatMassWilsonZTwoHeavy}{\LatMassWilson^{Z_2}_\text{heavy}}
\newcommand{\LatMassStaggered}{m}
\newcommand{\LatMassStaggeredTric}{\LatMassStaggered^\text{tricr.}}
\newcommand{\LatMassStaggeredTricHeavy}{\LatMassStaggeredTric_\text{heavy}}
\newcommand{\LatMassStaggeredTricLight}{\LatMassStaggeredTric_\text{light}}
\newcommand{\LatMassStaggeredTricHeavyPion}{\LatMassStaggeredTric_{\pi \text{ heavy}}}
\newcommand{\LatMassStaggeredTricLightPion}{\LatMassStaggeredTric_{\pi \text{light}}}
\newcommand{\LatCoupling}{\beta}
\newcommand{\LatCouplingC}{\LatCoupling_c}

\newcommand{\Poly}{L}
\newcommand{\PolyIm}{\Poly_\text{Im}}
\newcommand{\PolySq}{\left|\left|\Poly\right|\right|}

\newcommand{\Binder}{B_4}
\newcommand{\Skewness}{B_3}
\newcommand{\Temp}{T}
\newcommand{\MuI}{\mu_i}
\newcommand{\MuIRW}{\mu_i^{RW}}
\begin{document}
%
\selectlanguage{english}
\title{%
Updates on the Columbia plot and its extended/alternative \\ versions}
\author{%
\firstname{Francesca} \lastname{Cuteri}\inst{1}\fnsep\thanks{Speaker, \email{cuteri@th.physik.uni-frankfurt.de}} \and
\firstname{Christopher} \lastname{Czaban}\inst{1,2}\fnsep \and
\firstname{Owe} \lastname{Philipsen}\inst{1,2}\fnsep \and
\firstname{Alessandro}  \lastname{Sciarra}\inst{1}
}
\institute{%
        Institut f\"{u}r Theoretische Physik, Goethe-Universit\"{a}t Frankfurt\\
        Max-von-Laue-Str.\ 1, 60438 Frankfurt am Main, Germany
    \and
        John von Neumann Institute for Computing (NIC) GSI\\
        Planckstr.\ 1, 64291 Darmstadt, Germany
}
\abstract{%
  We report on the status of ongoing investigations aiming at locating the deconfinement critical point with standard Wilson fermions and ${\Nf=2}$ flavors towards the continuum limit (standard Columbia plot); locating the tricritical masses at imaginary chemical potential with unimproved staggered fermions at $\Nf=2$ (extended Columbia plot); identifying the order of the chiral phase transition at $\mu=0$ for $\Nf=2$ via extrapolation from non integer $\Nf$ (alternative Columbia plot).
}
\maketitle
\section{Introduction}\label{intro}

\begin{figure}[tp]
   \centering
   \subfigure[First order scenario in the $\ms-\mud$ plane]%
             {\label{fig:firstOrderScenarioCP}\includegraphics[width=0.45\textwidth,clip]{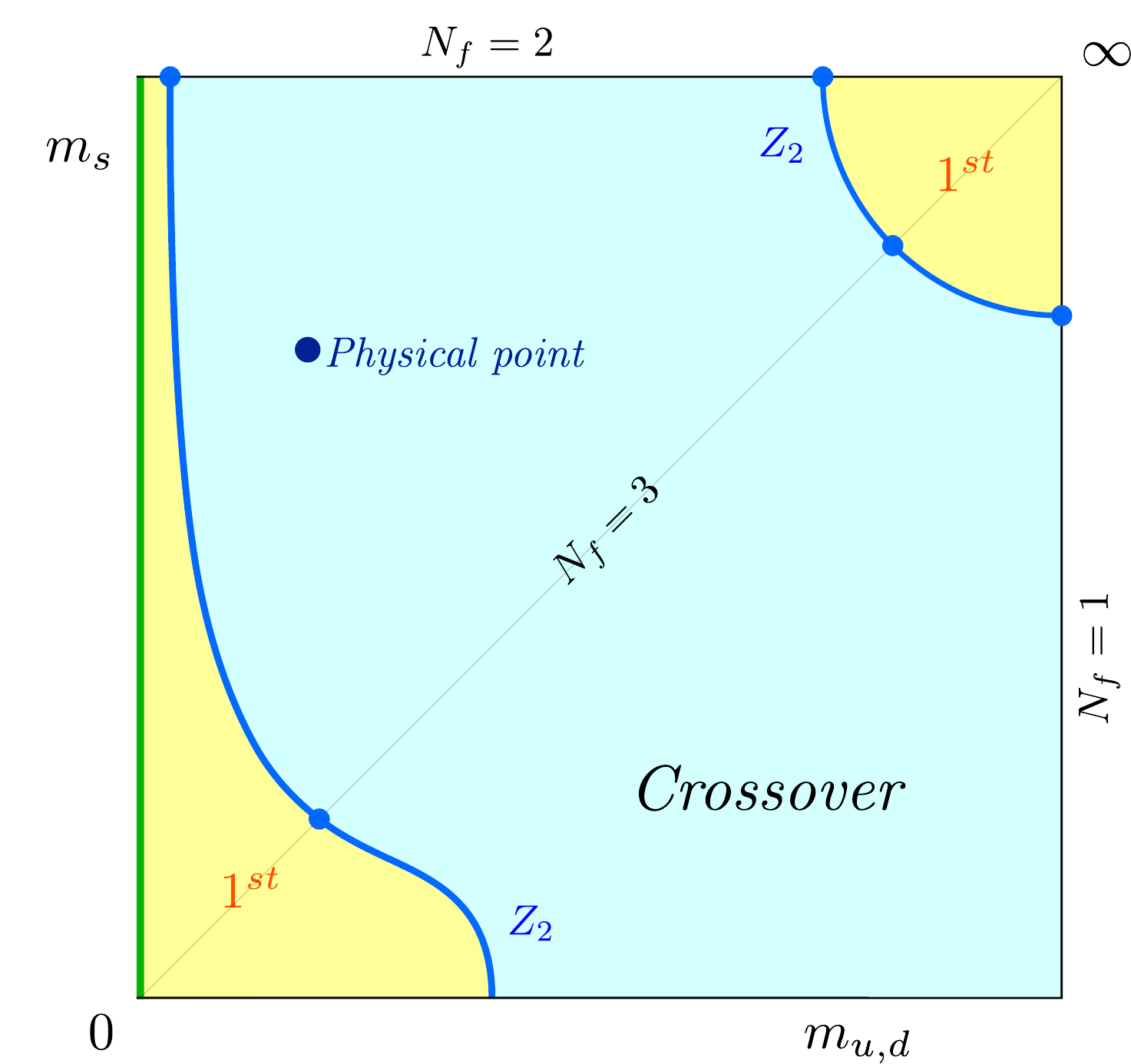}}\hfill
   \subfigure[Second order scenario in the $\ms-\mud$ plane.]%
             {\label{fig:secondOrderScenarioCP}\includegraphics[width=0.45\textwidth,clip]{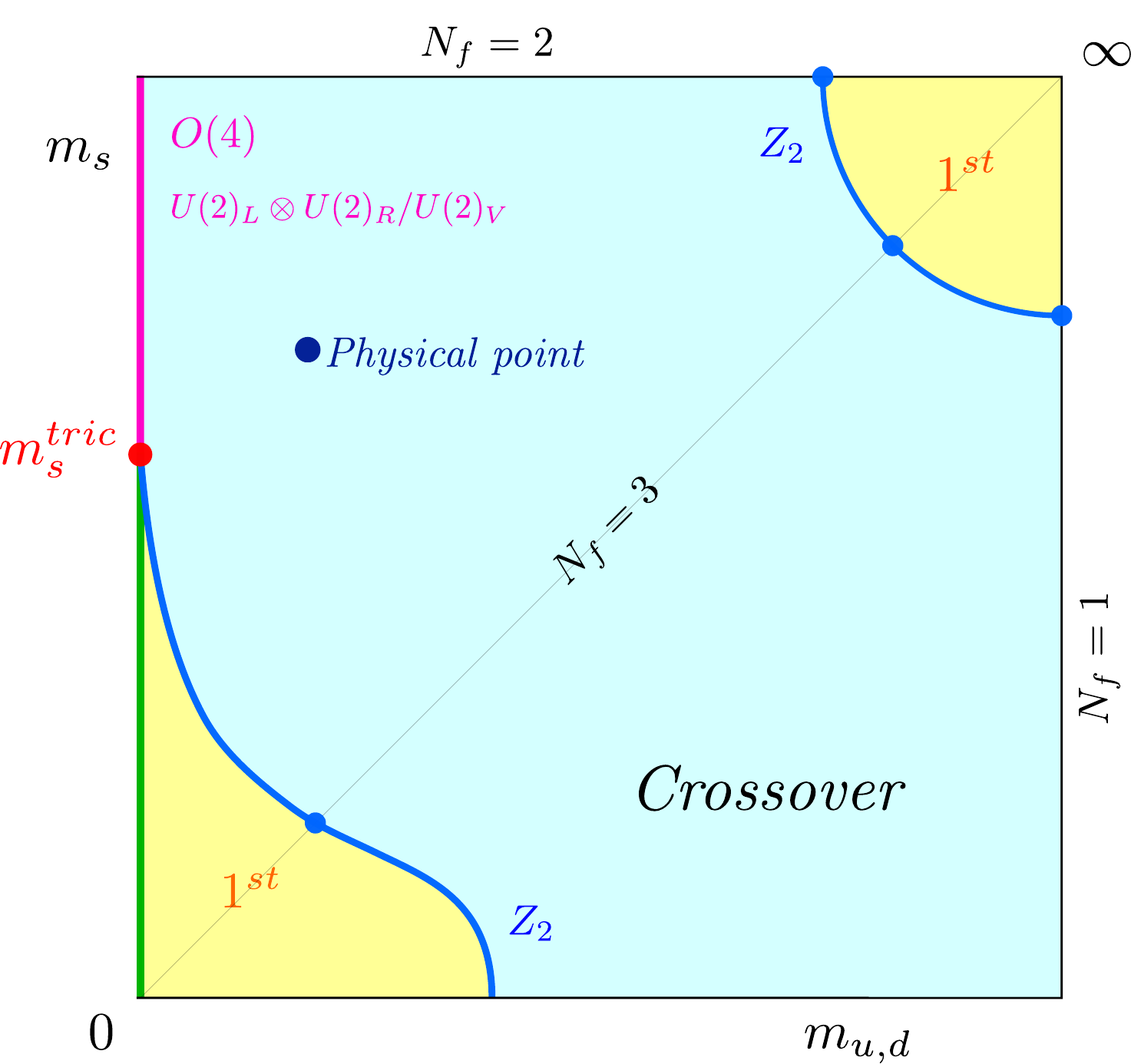}}
   \caption{Two possible scenarios for the order of the QCD thermal phase transition as a function of the quarks masses. Indicated in Fig.~\protect\ref{fig:secondOrderScenarioCP} are also plausible universality classes for the second order line at $\mud=0$}
   \label{fig:scenariosCP}
\end{figure}

Current findings and/or expectations of the lattice QCD community on the order of the thermal phase transition in QCD as function of the two light (assumed degenerate) quark masses $\mud$ and the strange quark mass $\ms$ are wrapped up in the \emph{Columbia plot} of which we show in \figurename~\ref{fig:scenariosCP} two possible versions in agreement with current findings, mostly not yet extrapolated to the continuum limit, findings.

The fact that the location of phase boundaries, is neither qualitatively (for the upper left corner), nor  quantitatively established, motivates us to push forward with more investigations on the subject aiming at clarifying, in particular, the picture for $\Nf=2$ degenerate light flavors. In this contribution updates are provided on our attempts to pursue the continuum limit location of the $Z_2$ critical endpoint in the heavy mass region $\LatMassWilsonZTwoHeavy$ at $\mu=0$ (\sectionname~\ref{sec-1}); use an extrapolation with tricritical exponents from non integer $\Nf$ for light quarks (\sectionname~\ref{sec-3}) and locate the tricritical heavy and light endpoints $\LatMassStaggeredTricHeavy$ and $\LatMassStaggeredTricLight$ at imaginary chemical potential (\sectionname~\ref{sec-2}). \sectionname~\ref{strategy} is instead devoted to a description of the strategy adopted in the various projects to locate the relevant phase transitions and establish their order.

\section{The (almost) common strategy}\label{strategy}

All numerical simulations have been performed using the publicly available~\cite{clqcd} \Ocl-based code \clqcd~\cite{Philipsen:2014mra}, which is optimized to run efficiently on \Amd  GPUs and provides, among others, an implementation of the (R)HMC algorithm for unimproved (rooted staggered) Wilson fermions. Moreover, all simulations were run on the \Loewe\ and \Lcsc\ clusters with the help of the Bash Handler to Monitor and Administrate Simulations (BaHaMAS) developed in the group~\cite{BaHaMAS}.

In our studies, the chemical potential has been kept fixed at either $\mu=0$ or at $\MuI=\MuIRW$, with the temperature $\Temp$ related to the coupling $\LatCoupling$ according to $\Temp=1/(\LatSpacing(\LatCoupling)\NTau)$. Studies in \sectionname~\ref{sec-2} and in \sectionname~\ref{sec-3} were conducted at a fixed temporal extent of the lattices (no continuum limit is attempted in these cases), while for the study described in \sectionname~\ref{sec-1} three different values of $\NTau$ were considered.
The ranges in mass $\LatMassStaggered$ or hopping parameter $\LatMassWilson$ and gauge coupling constant $\LatCoupling$ were always dictated by our purpose of locating the chiral/deconfinement phase transition with the purpose of mapping down the position and nature of critical boundaries in the phase diagram.
 
To locate the chiral/deconfinement phase transition and to identify its order, a "common" strategy was adopted, which consisted in a finite size scaling analysis (FSS) of the third and/or fourth standardized moments of the distribution of the (approximate) order parameter.
The $\text{n}^{\text{th}}$ standardized moment, given the distribution of a generic observable $\mathcal{O}$, is expressed as
    \begin{equation}
        B_n(\mathcal{O})=\frac{\left\langle\left(\mathcal{O} - \left\langle\mathcal{O}\right\rangle\right)^n\right\rangle}{\left\langle\left(\mathcal{O} - \left\langle\mathcal{O}\right\rangle\right)^2\right\rangle^{n/2}},
    \end{equation}
and we will analyze its dependence on some parameter $X\in\left\lbrace\LatMassStaggered,\LatMassWilson,\LatCoupling\right\rbrace$ and on the volume.
We will introduce the (approximate) order parameter $\mathcal{O}$ for each investigation in the corresponding sections. However, in all cases, in order to extract the order of the transition as a function of the bare quark mass and/or number of flavors, we considered the kurtosis $\Binder(\mathcal{O})$ of the sampled distribution of $\mathcal{O}$.

For the studies described in \sectionname~\ref{sec-1} and in \sectionname~\ref{sec-3} it was enough to consider the kurtosis evaluated on the phase boundary i.e. in correspondence to the value of the coupling $\LatCouplingC$ at which the distribution, at various volumes $\NSigma$ , showed a vanishing skewness $\Skewness(\LatCoupling=\LatCouplingC)=0$.
For the study described in \sectionname~\ref{sec-2}, the analysis is slightly different due to the more complex phase structure of QCD at imaginary chemical potential, that was first studied by Roberge and Weiss. The order parameter for the Roberge-Weiss phase transition shall be used in this case. Every coupling is, then, critical and the thermal phase transition is located by the position in $\LatCoupling$ of the crossing of the kurtosis datasets at various $\NSigma$ values.
%
%

In the thermodynamic limit $\NSigma \rightarrow \infty$, the universal values taken by the kurtosis $\Binder$ and by the critical exponent $\nu$ are well known results. 
However, the discontinuous step function characterizing the thermodynamic limit is smeared out to a smooth function as soon as a finite volume is considered and a FSS is needed. In all cases we varied the spatial extent of the lattice $\NSigma$ such that  the aspect ratios, governing the size of the box in physical units at finite temperature, was in the range $\NSigma/\NTau\in[2-5]$.
In the vicinity of a critical point, the kurtosis can be expanded in powers of the scaling variable $x=(X - X_{c}) \NSigma^{1/\nu}$ and, for large enough volumes, the expansion can be truncated after the linear term,
\begin{equation}\label{eq:BinderScaling}
    \Binder(\LatCouplingC, X, \NSigma) \simeq  \Binder(\LatCouplingC,X_{c}, \infty) + c (X - X_{c}) \NSigma^{1/\nu}.
\end{equation}
In our case, for the studies described in \sectionname~\ref{sec-1} and in \sectionname~\ref{sec-3}, the critical value for $X_{c}=\LatMassWilson,\LatMassStaggered$ corresponds to a second order phase transition in the 3D Ising universality class, 
so that one can fix $\Binder \approx 1.604$ and $\nu \approx 0.63$ and perform the fit to Eq.~\eqref{eq:BinderScaling} with the sole aim of extracting $X_c$ (and $c$).

\subsection{The quantitative data collapse as alternative to the kurtosis fit}\label{substrategy}

The technique of fitting the kurtosis around $X_c$ has been employed in many studies in order to locate a phase boundary and/or establish the order of a phase transition. However, especially when it comes to fitting reweighted, rather than just raw, data and due to the necessity of fixing different fit ranges for different volumes, the fit procedure ends up relying on some more or less arbitrary decisions hence being not so solid.
Based on these observation an alternative procedure was devised and implemented, which measures in a quantitative and solid way the quality of the collapse among sets of data obtained on different lattice sizes once they are plotted against the scaling variable $x$, with the critical parameter $X_{c}$ and the exponent $\nu$ suitably fixed. More details on this method can be found in~\cite{SciarraThesis}.

The \emph{collapse quality}, fixed some critical $\bar{X}_c$ value and some value $\bar{\nu}$ for the critical exponent and considering all pairs of volumes, is
\begin{equation}
 Q(\bar{X}_c,\bar{\nu}) \equiv \frac{1}{\Delta x} \int^{x_{\text{max}}}_{x_{\text{min}}} \frac{1}{N_V} \sum_{i=1}^{N_V} \sum_{j=1}^{N_V} \Theta (j-i) \left[\Binder\left(x(\bar{X}_c,\bar{\nu},V_i)\right) - \Binder\left(x(\bar{X}_c,\bar{\nu},V_j)\right)\right]^2 \mathrm{d}x,
\end{equation}
where $N_V$ is the number of simulated volumes, the integration is done numerically (after having reweighted with a high enough resolution to safely interpolate) and $\Delta x=x_{\text{max}}-x_{\text{min}}$ is the symmetric interval around $x=0$ over which the integration takes place. Once the interpolation and the integration are made, $Q(X_c,\nu)$ is minimized as a function of its two variables. Since a procedure to get the statistical errors on $X_{c}$ and $\nu$ needs to be also set up, what one can do is to use $N_{\text{boot}}$ different sets of reweighted kurtosis to minimize $Q(\bar{X}_c,\bar{\nu})$, so that as many different estimates of $X_{c}$ and $\nu$ are obtained and the corresponding statistical error can be determined.
Clearly, a role is also played by the choice of $\Delta x$. However, while using a too wide $\Delta x$ can be wrong due to the critical region being smaller, one can extrapolate $X_{c}$ and $\nu$ to $\Delta x \to 0$.

Contrary to the kurtosis fit procedure, in the quantitative data collapse reweighting with a higher resolution can only help getting a better estimate of $Q(\bar{X}_c,\bar{\nu})$~\footnote{Although the result of the interpolation becomes soon very stable against the number of reweighted data.} and the arbitrariness in the choice of $\Delta x$ is removed by the extrapolation. For the above reasons the quantitative data collapse should be used, if not to replace, at least to cross check results from the kurtosis fit.
\clearpage
\section{Updates on the Columbia plot: $Z_2$ boundary in in the high masses corner}\label{sec-1}

\begin{figure}[tp]
   \centering
   \subfigure[Shift of the critical $m_{Z_2}$ masses at $\Nf=2$ towards the continuum limit]%
             {\label{fig:Nf2ContinuumLimitHighMassesHighlighted}\includegraphics[width=0.38\textwidth,clip]{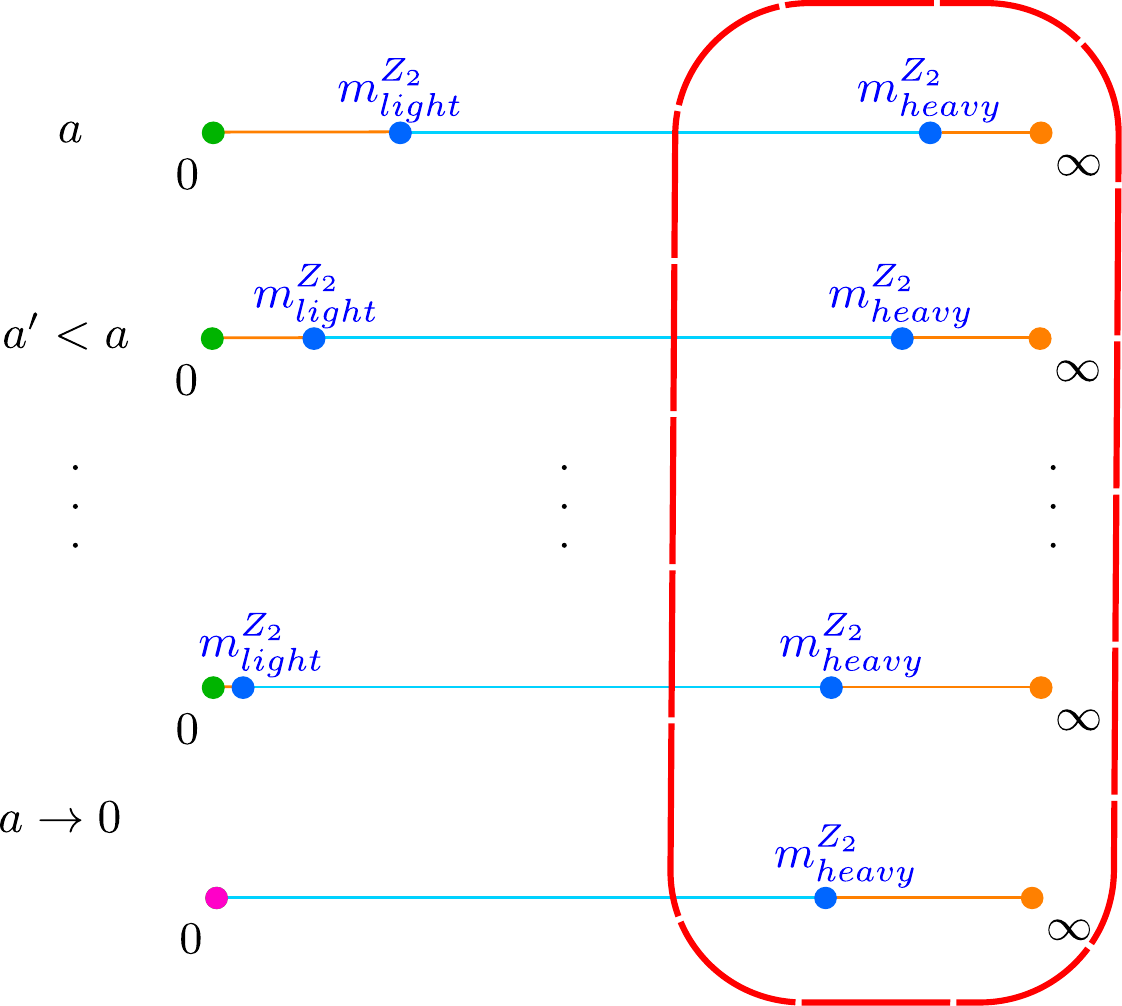}}\hfill
   \subfigure[FSS of $\Binder$ and fit]
             {\label{fig:binderFitHighMasses}\includegraphics[width=0.48\textwidth,clip]{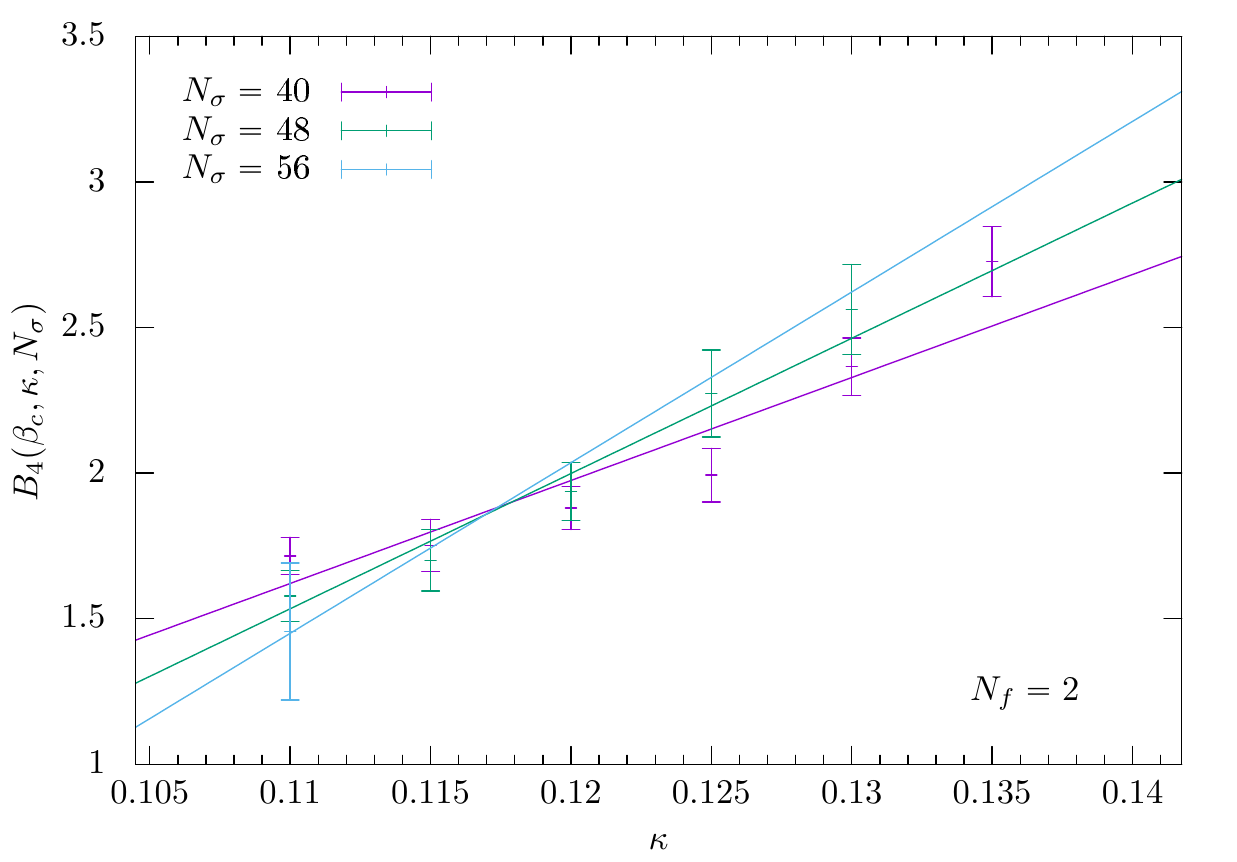}}
   \caption{Features and preliminary results on the $Z_2$ boundary in the high-mass(es) corner of the Columbia plot.}
   \label{fig:highMasses}
\end{figure}
\renewcommand{\arraystretch}{1.2}
\begin{table}[b]\scriptsize\centering
    \begin{tabular}{ccccccc}
        \toprule[0.3mm]
        $\NTau$ & $\LatMassWilsonZTwoHeavy$ & $a$ [fm] & $a m_\pi$ & $m_\pi$ [MeV] & $V_{min}$ [fm$^3$] & $L_{min}$ [fm]\\
        \midrule[0.1mm]
        6  & 0.0890(27) & [0.118(1):0.123(1)] & [3.108:2.241] & [5198(35):3589(15)] & 50 & 3.68 \\ 
        8  & 0.1128(29) & [0.088(1):0.092(1)] & [2.131(1):1.397(1)] & [4768(22):2995(33)] & 45 & 3.56 \\
        \bottomrule[0.3mm]
    \end{tabular}
    \caption{Results on $V_{\text{min}}$ from fits of the kurtosis.}
    \label{tab:minimalPhysicalVolume}
\end{table}
The entity of the cut-off effects that quantitatively affect our picture of the QCD phase structure have been investigated in previous studies in the upper right corner of the Columbia plot and the $Z_2$ transitions were already observed to shift to smaller masses for $\Nf=2, 2+1, 3$ at $\mu=0$ \cite{deForcrand:2008zi,deForcrand:2007rq,Saito:2011fs,Fromm:2011qi}. A sketch of this behavior for $\Nf=2$ is given in \figurename~\ref{fig:Nf2ContinuumLimitHighMassesHighlighted}. No continuum limit result is known yet.

In this case the norm of the Polyakov loop $\PolySq$ was used as approximate order parameter for the deconfinement phase transition. The case of $\Nf=2$ degenerate quarks was addressed and, with a scan in $\LatMassWilson$, the location of the critical $\LatMassWilsonZTwoHeavy$ endpoint on $\NTau=6,8,10$ was investigated with the aim to monitor and possibly model cut-off effects. A previous report on this project is to be found in~\cite{Czaban:2016yae}, while here we would like to just focus on the impact of finite size effects in this investigation. It is surely a feature of the heavy mass region that pions, with the adopted discretization, cannot be resolved on our lattices up to $\NTau~\approx~10$. And, while $\LatSpacing \to 0$ with growing $\NTau$, the necessity of keeping the relation $1 \ll \NTau \ll \NSigma$ satisfied forces us to use larger $\NSigma$ values to keep the size of the box fixed. This has to be true already for the smallest of the volumes in our FSS analysis.

In view of the increasing cost of simulations some work was invested in devising an alternative and possibly cheaper strategy to locate $\LatMassWilsonZTwoHeavy$. One could, indeed, first identify at any fixed value of the bare mass some \emph{minimal physical volume} $V_{\text{min}}$ characterized by allowing a reliable extraction of $\LatMassWilsonZTwoHeavy$ out of a linear fit of the kurtosis. At a different $\LatMassWilson$ or $\NTau$, it should be then enough to e.g. reweight the effective potential $V_{\text{eff}}$ at just one fixed $V\gtrsim V_{\text{min}}$ as in~\cite{Saito:2011fs} to locate the phase transition and understand its nature.
In practice we start by using a modified fit ansatz for the kurtosis in the vicinity of the critical point~\cite{Jin:2017jjp}
\begin{equation}
    \Binder(\LatMassWilson, N_{\sigma}) = \left[\Binder(\LatMassWilsonZTwoHeavy,\infty) + \textcolor{blue}{c}\,(\LatMassWilson - \textcolor{blue}{\LatMassWilsonZTwoHeavy}) \NSigma^{(1/\nu)}\right](1 + \textcolor{blue}{B}\NSigma^{y_t-y_h})\nonumber,
\end{equation}
which incorporates the finite volume effect for generic observables which are a mixture of energy-like and magnetization-like operator and where the value of the exponent $y_t-y_h$ is fixed by universality. Then we estimate $V_{\text{min}}$ by excluding one-by-one the smallest physical volumes in the fit, until the value of the coefficient of the correction term is compatible with zero. Preliminary results are collected in \tablename~\ref{tab:minimalPhysicalVolume} and  one example of the performed fits is provided in \figurename~\ref{fig:binderFitHighMasses}.

\section{Updates on an alternative Columbia plot: $Z_2$ boundary at non-integer $\Nf$}\label{sec-3}
\begin{figure}[tp]
   \centering
   \subfigure[First order scenario in the $\mud-\Nf$ plane]%
             {\label{fig:firstOrderScenarioMvsNf}\includegraphics[width=0.45\textwidth,clip]{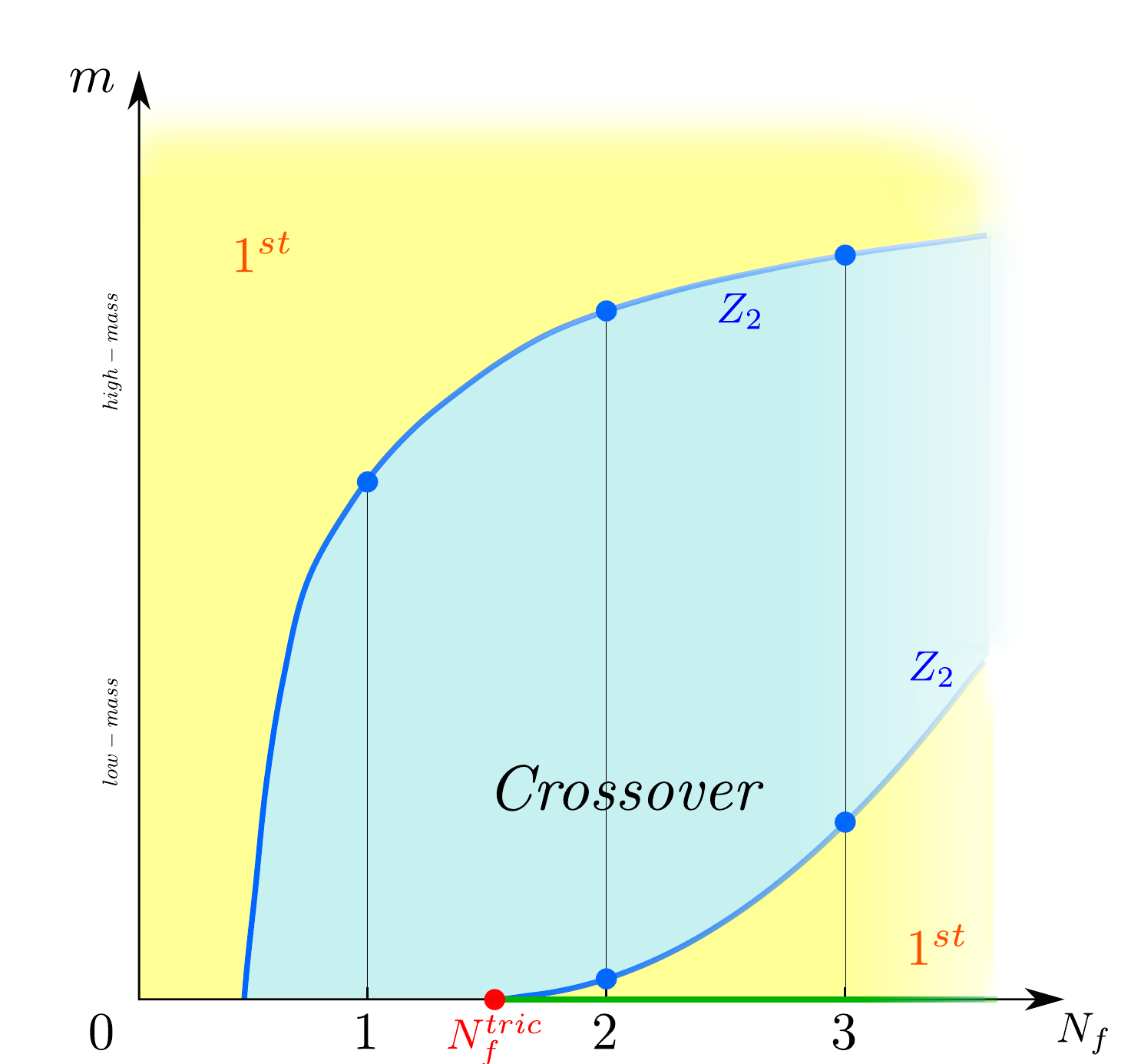}}\hfill
   \subfigure[Second order scenario in the $\mud-\Nf$ plane]%
             {\label{fig:secondOrderScenariosMvsNf}\includegraphics[width=0.45\textwidth,clip]{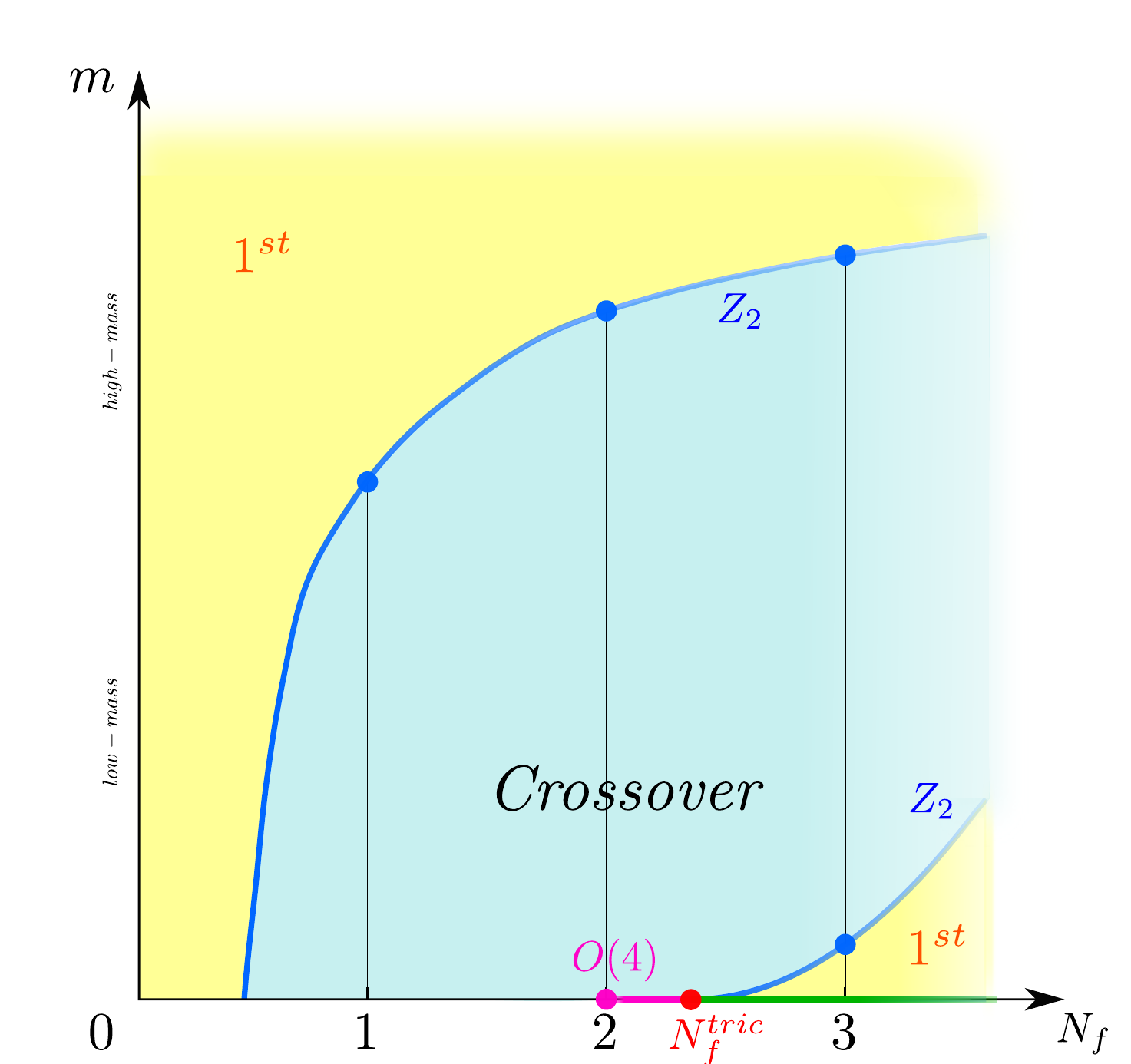}}
   \caption{The two considered possible scenarios for the order of the QCD thermal phase transition as a function of the light-quarks mass and the number of fermion flavors.}
   \label{fig:scenariosMvsNf}
\end{figure}
In this study, we consider QCD with $\Nf$ mass-degenerate quarks of mass $\LatMassStaggered$ at zero density and the partition function reads 
\begin{equation}
    Z_{\Nf}(m)=\int \mathcal{D}U \left[\det M(U,m)\right]^{\Nf} e^{-\SGluon}\;.
\end{equation}
We formally view this as a partition function of some statistical system characterized by a continuous parameter $\Nf$ and we try to find out for which (tricritical) value of $\Nf$ the phase transition displayed by this system changes from first-order to second-order.
Of course, the extension of $Z_{\Nf}(m)$ to non-integer values of $\Nf$ is not unique, and since we use an interpolation characterized by non-integer powers of the determinant, it does not correspond to any local quantum field theory.
However, we are just considering, for any given lattice spacing, a statistical system that represents one 
particular interpolation between two quantum field theories with integer $\Nf$.
It is not the specific value of $\Nf^{tric}$ being relevant, but just its relative location with respect to the integer $\Nf=2$ (see Figure~\ref{fig:scenariosMvsNf}). Such result should not depend on the chosen interpolation.

In terms of discretization, we employ staggered fermions, where the first-order region is narrow already on coarse lattices. The RHMC algorithm~\cite{Kennedy:1998cu} can be used to simulate any number $\Nf$ of degenerate flavors of staggered fermions, with $\frac{\Nf}{4}$ being the power to which the fermion determinant is raised in the lattice partition function.
%

\begin{wrapfigure}{r}{0.5\textwidth}
   \centering
              {\label{fig:rescaled}\includegraphics[width=0.5\textwidth,clip]{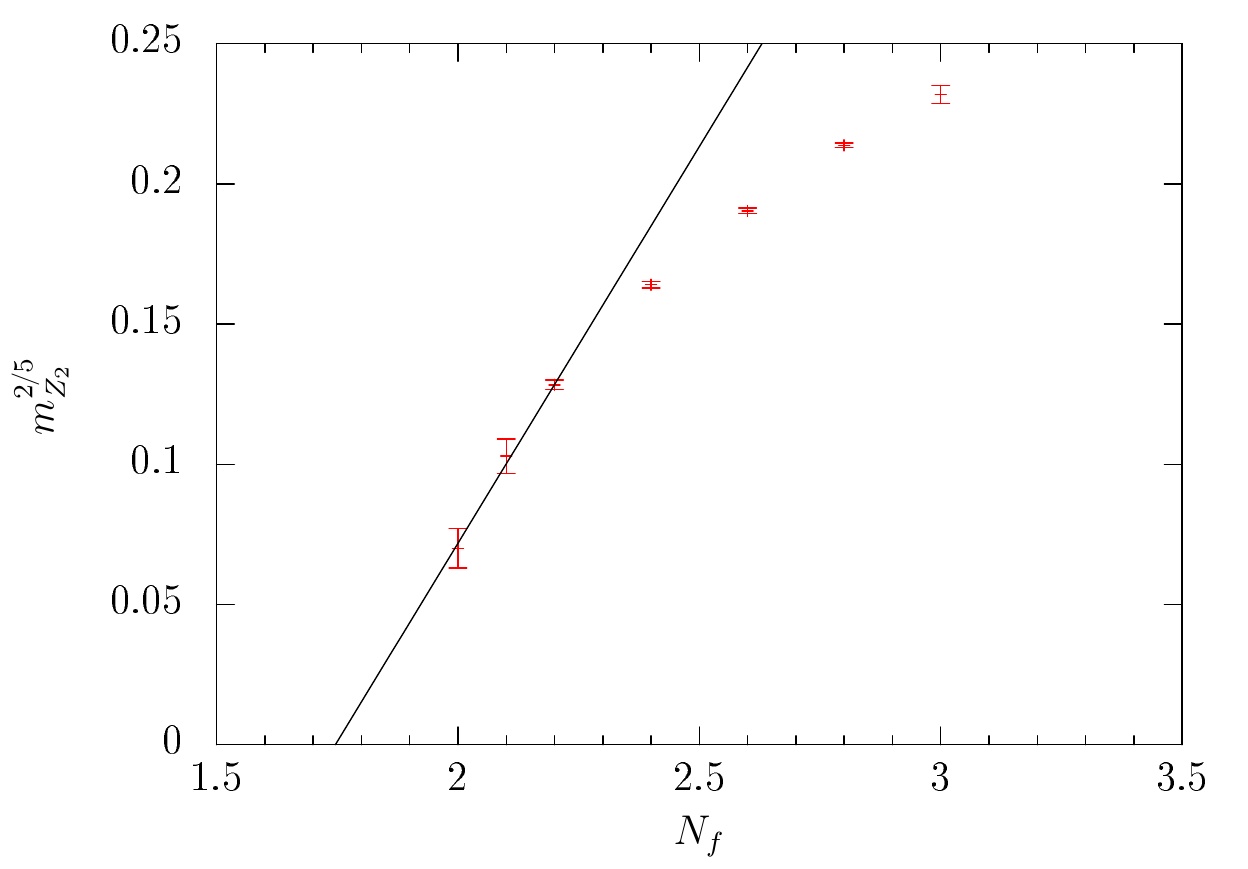}}
   \caption{$Z_2$ critical line in the $m^{2/5}-\Nf$ plane. The solid line is a linear fit in the range $[2.0, 2.2]$. For $\Nf=2.0$ we use the result from~\protect\cite{Bonati:2014kpa}.\label{fig:tricrScaling}}
\end{wrapfigure}
We measured the chiral condensate $\chiralcond$ as approximate order parameter at $5$ different $\Nf$ values (${\Nf=2.8, 2.6, 2.4, 2.2, 2.1}$). For each parameter set $\lbrace \Nf, m, \NSigma, \LatCoupling \rbrace$, statistics of about $(200k-400k)$ trajectories were accumulated over $4$ Markov chains, subject to the requirement that the skewness of the chiral condensate distribution is compatible within $\sim2-3$ standard deviations among the different chains.
The phase transition was located according to the kurtosis fit analysis described in \sectionname~\ref{strategy}. 
We observed how, as $\Nf$ is lowered, $m_{Z_2}$ decreases towards zero and we used that the critical line, in the vicinity of a tricritical point, is known to display a power law dependence with known critical exponents~\cite{lawrie}.
The scaling law is 
\begin{equation}
    m^{2/5}_{Z_2}(\Nf)=\textcolor{blue}{C}\left(\Nf - \textcolor{blue}{\Nf^{tric}}\right). 
\end{equation}
The plot of the rescaled mass $m^{2/5}_{Z_2}$ as a function of $\Nf$ is displayed in Figure~\ref{fig:tricrScaling}. The solid line in the mass-rescaled plot shows that the simulated results for $\Nf=2.2,2.1$ are aligned with the result for $\Nf=2.0$, which we were also able to roughly cross-check via direct simulations, of the extrapolation from imaginary chemical potential from~\cite{Bonati:2014kpa}. This preliminarily indicates agreement with the expected scaling relation for $\Nf\le2.2$.
%

\section{Updates on the extended Columbia plot: Roberge-Weiss endpoint}\label{sec-2}
\begin{figure}[tp]
   \centering
   \subfigure[The Roberge-Weiss Columbia plot i.e. the order of the thermal phase transition in the $\ms-\mud$ plane at $\MuI=\MuIRW$.The $\Nf=2$ case is highlighted.]%
             {\label{fig:robergeWeissColumbiaPlotNf2Highlighted}\includegraphics[width=0.36\textwidth,clip]{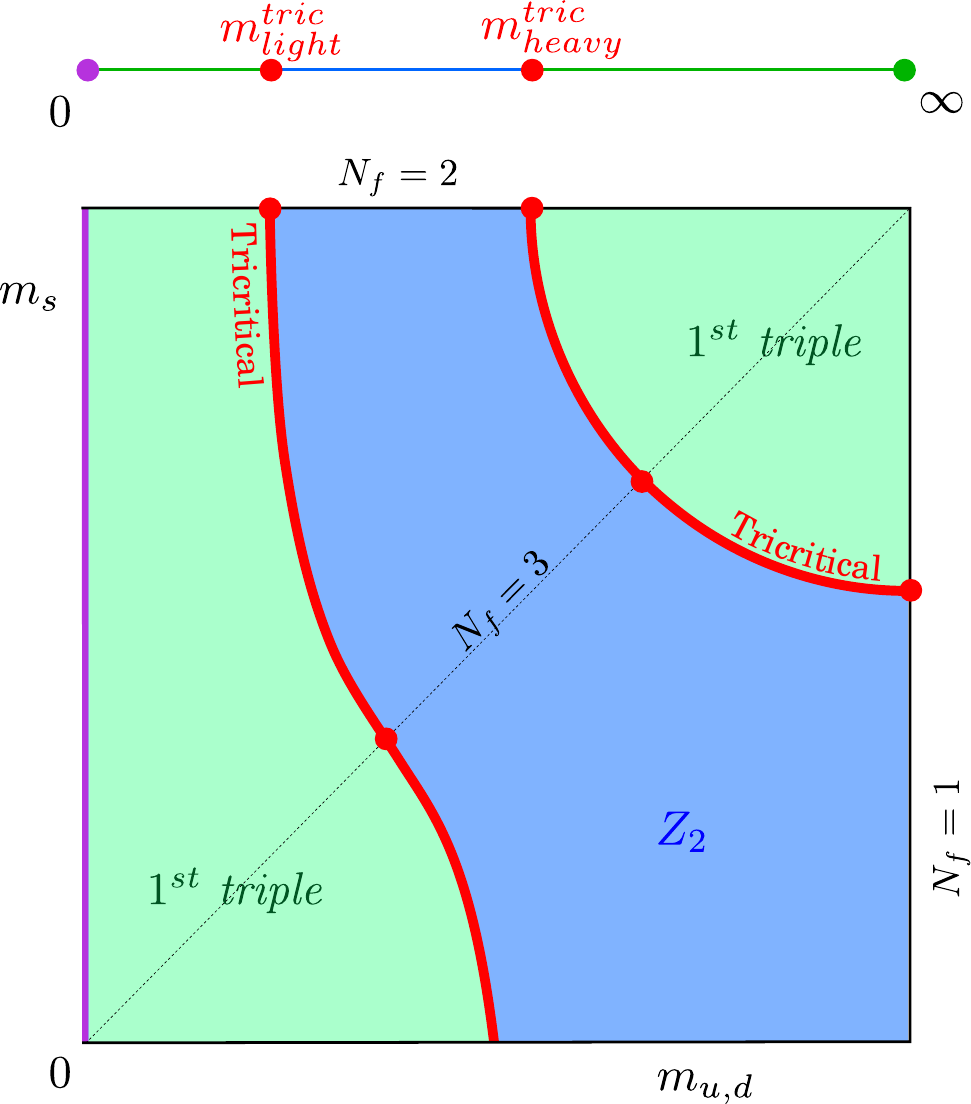}}\hfill
   \subfigure[Critical exponent $\nu$ as a function of the bare quark mass $\LatMassStaggered$. For the sake of readability, the mass axis has been broken and two different scales have been used.  Shaded points correspond to preliminary results for which higher statistics is needed.]%
             {\label{fig:NuVSmass}\includegraphics[width=0.59\textwidth,clip]{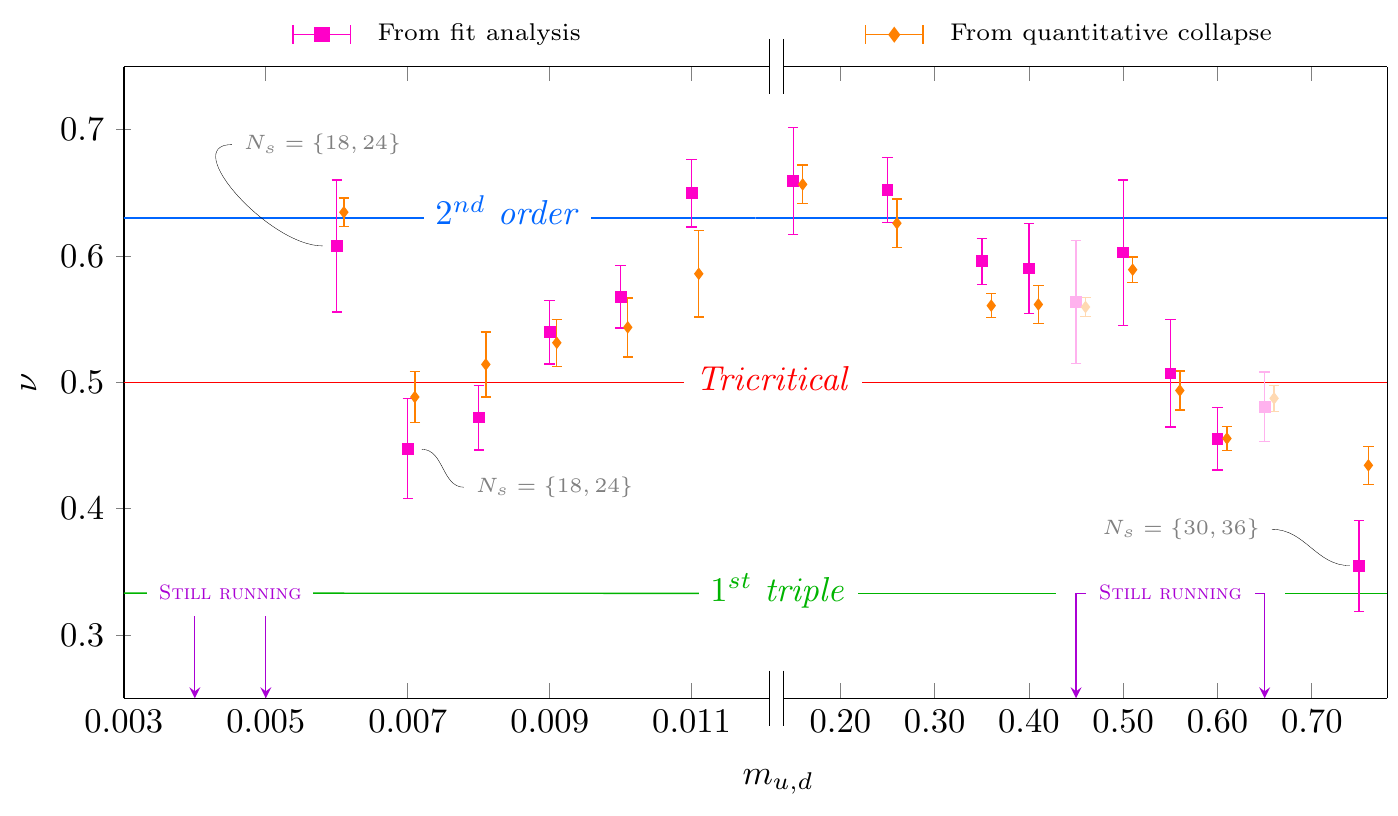}}\hfill
   \caption{Features and preliminary results on the tricritical endpoints in the Roberge-Weiss Columbia plot}
   \label{fig:staggeredRW}
\end{figure}

The Columbia plot at $\MuI=\MuIRW$ displayed in \figurename~\ref{fig:robergeWeissColumbiaPlotNf2Highlighted} looks similar to the one in \figurename~\ref{fig:firstOrderScenarioCP}, but with the $Z_2$ lines replaced by tricritical lines, first order triple regions that are wider than at $\mu=0$ and a second order $Z_2$ region at intermediate values for the quark masses. In this case the imaginary part of the Polyakov loop $\PolyIm$ was measured as order parameter for the Roberge-Weiss phase transition. Once again, we focused on the case of $\Nf=2$ degenerate unimproved staggered quarks, and tried to locate, with a scan in mass, the tricritical points $\LatMassStaggeredTricHeavy$ and $\LatMassStaggeredTricLight$ on $\NTau=6$ lattices as already done for other discretizations and $\NTau$ values~\cite{Philipsen:2014rpa,Cuteri:2015qkq,PhysRevD.83.054505}. A previous report on this project is to be found in~\cite{Philipsen:2016swy}.

For each value of $\mud$, simulations were performed at a fixed temporal lattice extent $\NTau=6$ and at a fixed value of the chemical potential $\LatSpacing\MuIRW=\pi/6$. The extraction of the critical exponent $\nu$ was accomplished both with the kurtosis fit procedure and with the quantitative data collapse described in \sectionname~\ref{substrategy}. Results for the critical exponent $\nu$ are reported in \figurename~\ref{fig:NuVSmass}. Since results from either kind of analysis happen to agree within a $1\sigma$ discrepancy in all (but one) case, they are combined to obtain the final answer on $\nu$.
To comment more on our results, it is important to stress that in the FSS, at least three simulated volumes (the largest) should be used. A third volume is essential to cross-check the position of the crossing of the kurtosis of different data sets in correspondence to the critical coupling $\LatCouplingC$. For each pair of points in \figurename~\ref{fig:NuVSmass} labeled by some indication on two (rather than three) volumes employed, simulations on a third larger volume are ongoing. For all other masses always three volumes have been simulated with $\NSigma^{\text{min}}=\ 12,18,24$ and $\NSigma^{\text{max}}=30,36,42$, depending on the mass. For each lattice size, 3 to 8 values of $\LatCoupling$ around the critical temperature were simulated, each with 4 Markov chains.
\begin{wrapfigure}{r}{0.5\textwidth}
   \centering
              {\includegraphics[width=0.5\textwidth,clip]{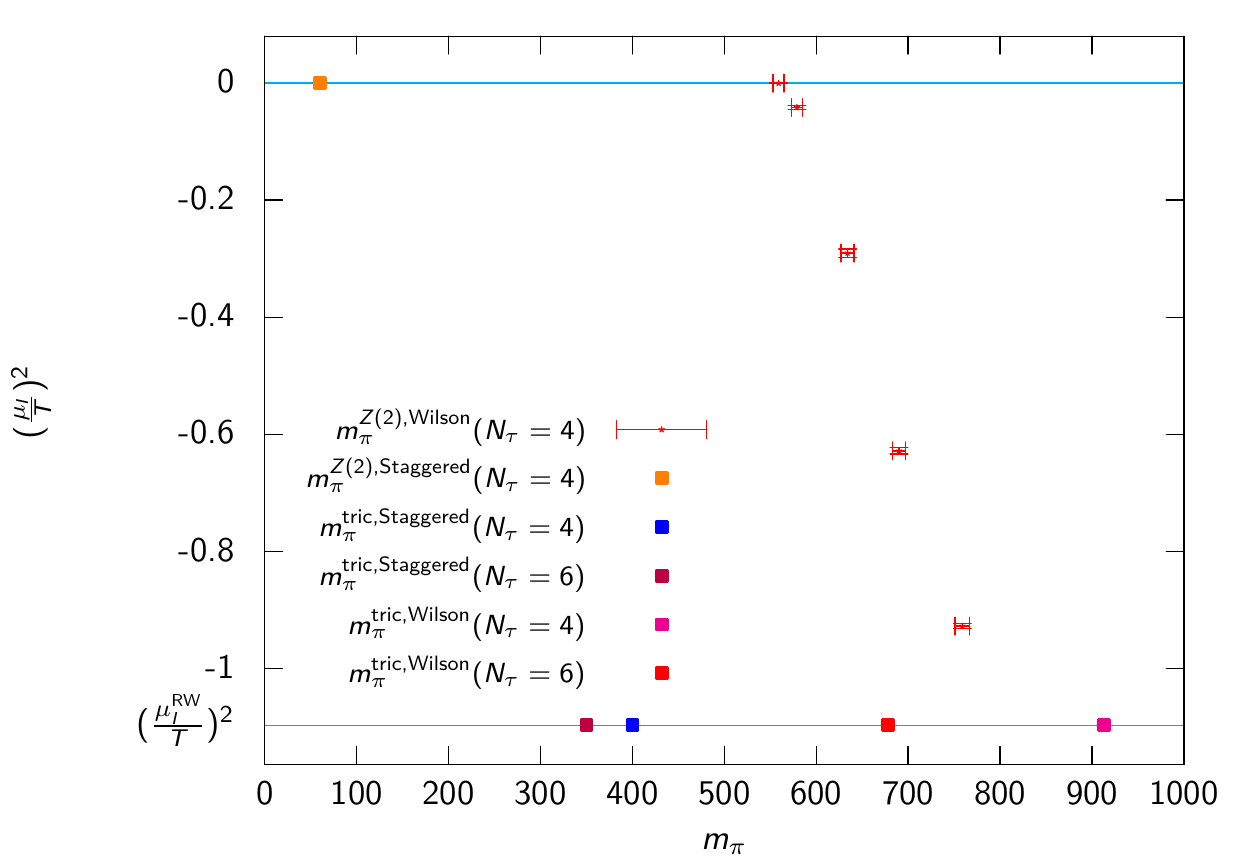}}
   \caption{Collection of results on the $Z_2$ critical line/points in the $\mpi-\mu^2$ plane. Results from~\cite{Bonati:2010gi,Bonati:2014kpa,Philipsen:2014rpa,Cuteri:2015qkq,Philipsen:2016hkv} are included. Various discretizations and/or results at different $\NTau$ can be compared.}\label{fig:nicePlot}
\end{wrapfigure}
In order to decide when to stop accumulating statistics, for large (small) masses the kurtosis of the imaginary part of Polyakov loop was required to be compatible on all the chains within 2 (3) standard deviations. Since this condition can be fulfilled also at a very poor statistics, due to large errors, a further empirical requirement is that values of the kurtosis from different chains must span, errors included, an interval not wider than $0.5$.

As indicated in \figurename~\ref{fig:NuVSmass} simulations for a few masses are still running and, at the moment, it is still not possible to give an indication on the position of the two tricritical masses with their statistical error, due to none of the simulated masses falling on the first order triple line.
What can be observed, e.g. by the shift in the crossing of subsequent pairs of kurtosis datasets, is that in the heavy (light) mass region larger and larger volumes are needed while the bare mass is increased (decreased) in ranges where the transition becomes tricritical to weak first order.
For this reason the preliminary estimate for the location of the two tricritical endpoints in terms of pion masses is unchanged with respect to those indicated in~\cite{Philipsen:2016swy}.
Here, we can, however, quote a preliminary indication, without errors, in terms of pion masses as $\LatMassStaggeredTricHeavyPion=2.809 \text{ GeV }$ and $\LatMassStaggeredTricLightPion=350 \text{ MeV }$~\cite{SciarraThesis}.
One can use this preliminary results to compare with results from other discretizations and/or at other $\NTau$ values. The comparison at small masses can be visualized by adding one more point to Fig. 6 in~\cite{Philipsen:2016hkv}, as we do in \figurename~\ref{fig:nicePlot}. It is possible, to e.g. consider the Wilson \emph{versus} staggered discretizations and compare the shift of $\LatMassStaggeredTricLightPion$ towards smaller masses going from $\NTau=4$ to $\NTau=6$: this is found to amount to $35\% (14\%)$ of the value for Wilson (staggered). At large masses the value found for $\LatMassStaggeredTricHeavyPion$ is, instead, unfortunately still affected by large cut-off effects ($\LatSpacing\mpi$ being still larger than 1).

\section{Acknowledgements}
This work is supported by the Helmholtz International Center for FAIR within the LOEWE program of the State of Hesse. We thank the staff of \Loewe\ for computer time and support. F.C. and O.P. are supported by the German BMBF under contract no. 05P1RFCA1/05P2015 (BMBF-FSP 202).

\end{document}